\documentclass[journal, draftcls, one column, 12pt]{IEEEtran}

\usepackage{wrapfig}

\usepackage[utf8]{inputenc} % allow utf-8 input
\usepackage[T1]{fontenc}    % use 8-bit T1 fonts
\usepackage{hyperref}       % hyperlinks
\usepackage{url}            % simple URL typesetting
\usepackage{booktabs}       % professional-quality tables
\usepackage{amsfonts}       % blackboard math symbols
\usepackage{nicefrac}       % compact symbols for 1/2, etc.
\usepackage{microtype}      % microtypography
\usepackage{mathtools}
\usepackage{color,multirow, multicol}
\usepackage[noend]{algpseudocode}

\newtheorem{proposition}{Proposition}
\usepackage{cite}
\usepackage{booktabs}
\usepackage{amsmath,amssymb,amsfonts}
\usepackage{graphicx}
\usepackage{textcomp}
\usepackage{xcolor}
\usepackage{braket}
\def\BibTeX{{\rm B\kern-.05em{\sc i\kern-.025em b}\kern-.08em
		T\kern-.1667em\lower.7ex\hbox{E}\kern-.125emX}}
\begin{document}
	
	\title{ Joint Concordance Index
	}

\author{Kartik~Ahuja and Mihaela~van der Schaar,~\IEEEmembership{Fellow,~IEEE}% <-this % stops a 
		\thanks{K. Ahuja and Mihaela van der Schaar are with the Department
		of Electrical and Computer Engineering, University of California, Los Angeles,
		CA, 90095 USA e-mail: (see ahujak@ucla.edu, mihaela@ee.ucla.edu)}
}
	\maketitle
	
	\begin{abstract}
		Existing metrics in competing risks survival analysis such as concordance and accuracy do not evaluate a model's ability to \textit{jointly} predict the event type and the event time. To address these limitations, we propose a new metric, which we call the joint concordance.  The joint concordance measures a model's ability to predict the overall risk profile, i.e., risk of death from different event types.
		We develop a consistent estimator for the new metric that accounts for the censoring bias. We use the new metric to develop a variable importance ranking approach.   Using the real and synthetic data experiments, we show that  models selected using the existing metrics are worse than those selected using joint concordance at jointly predicting the event type and event time. We show that the existing approaches for variable importance ranking often fail to recognize the importance of the event-specific risk factors, whereas, the proposed approach does not, since it compares risk factors based on their contribution to the prediction of the different event-types. 
		To summarize, joint concordance is helpful for model comparisons and variable importance ranking and has the potential to impact applications such as risk-stratification and treatment planning in  multimorbid populations. 
	\end{abstract}
	\section{Introduction}
	
	\subsection{Motivation}
	\label{motivation_section}
	The concordance index [1] is one of the most widely used metrics in survival analysis with competing risks (SA-CR) for measuring a model's discriminative ability, i.e., a model's ability to correctly order the subjects based on their risk. As was pointed out in [1], the concordance index is used to assess the prognostic ability of a model for one event type of interest in the presence of competing risks,  but it is not adequate to assess the prognostic ability of a model when there is more than one event type of interest.  Several clinical scenarios  with more than one event types are given below. 
	\begin{enumerate}
		\item  \textbf{Treatment planning for multimorbid populations.} Adverse treatment reactions are one of the leading causes of death  in the United States [2]. This has generated a large amount of interest to tackle the problem of polypharmacy  [2].  Multi-morbidity $-$ the accumulation of chronic diseases $-$ has emerged as a major contemporary
		challenge of the ageing population [3]. More than two-thirds of people aged over 65 are multimorbid,
		i.e., have two or more chronic diseases [4]. The current healthcare provision is not designed to consider diseases in
		combination leading to complications arising from unnecessary polypharmacy. Therefore, it is important to develop treatment plans in multimorbid populations after assessing the overall risk profile, i.e. the risks of death from different conditions. 
		
		\item \textbf{Treatment planning for critical care.} SA-CR models have been used to develop early warning systems to predict the event time and the event type (e.g., ventilation or discharged alive, different types of organ failures) [5]. In these applications, the joint prediction of the event type (e.g., ventilation) and the event time is helpful for planning the allocation of resources (e.g., ventilator). 
	\end{enumerate}
	We now elaborate on why the existing metrics are not sufficient to cater to the above scenarios.

	\begin{figure}
		\begin{center}
			\includegraphics[trim= 0mm 40mm 20mm 0mm, width=5in]{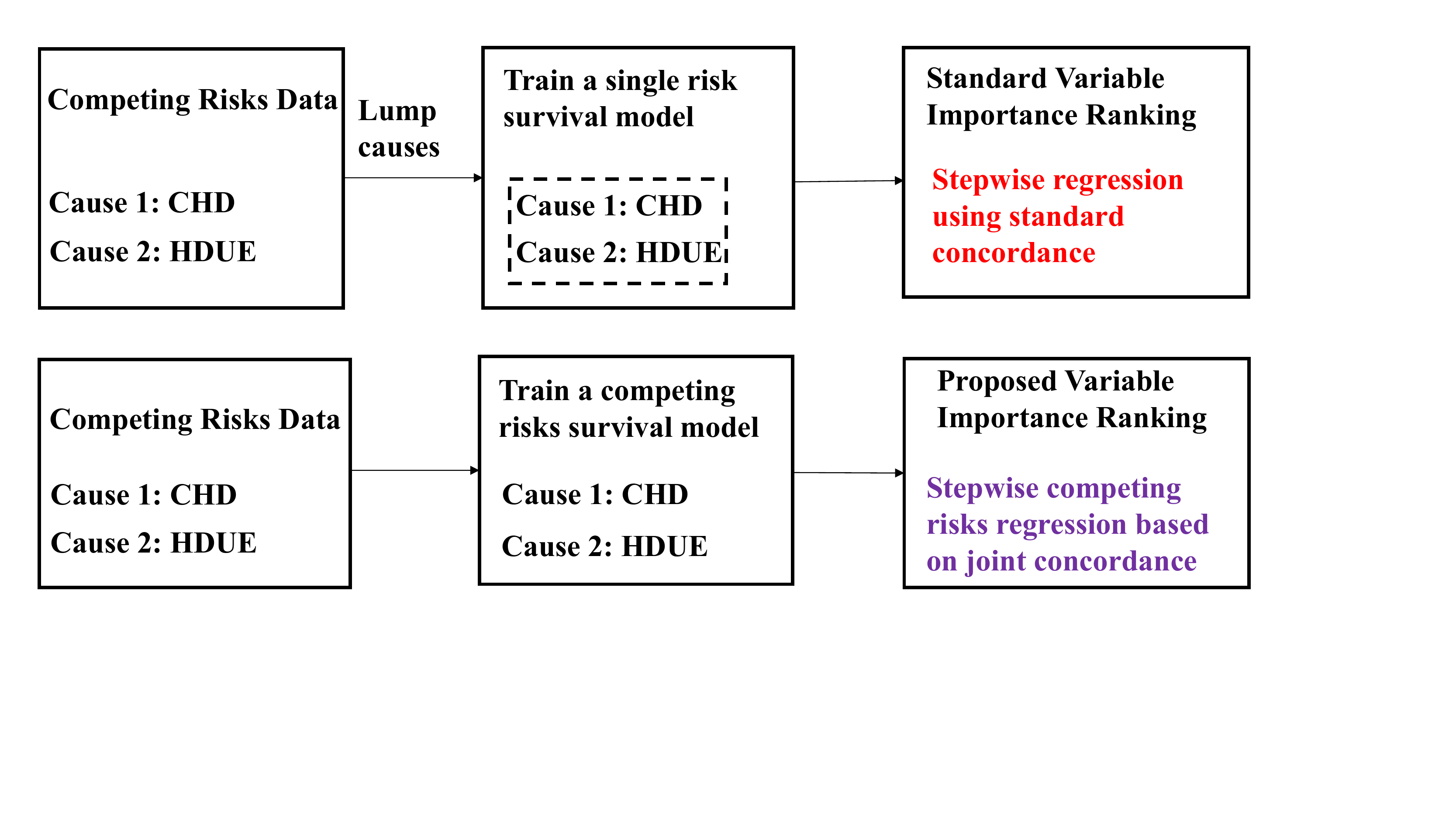}
			\caption{Variable importance ranking for overall risk profile: standard vs proposed; CHD: Coronary Heart Disease, HDUE: Heart Disease of Uncertain Etiology}
			\label{figure1}
		\end{center}
	\end{figure}

	\subsection{Contributions}
	
	In this work, we propose a new metric that we call the \textit{joint concordance}.   The joint concordance index is the probability that a given model  accurately predicts the event type for a subject while also ranking that subject's risk  correctly among the other subjects.  We show that the joint concordance index can be interpreted by decomposing it into a metric that is similar to accuracy and  concordance conditional on the correct predictions. We also prove that the decomposition does not lead to standard metrics, i.e., the concordance and accuracy. In addition, we prove that the joint concordance cannot be expressed as a function of the concordance and the accuracy.

	In most survival analysis settings, the most common form of  censoring that is observed is right censoring; right censoring occurs when either the subject is lost in the follow-up, or the study ends. In these scenarios, the estimation of the joint concordance index can become more difficult as censoring can introduce bias in the population that is observed at different times in the follow-up.  Different models can be used to estimate the censoring distribution such as the Kaplan-Meier model, the Cox model. We estimate the model of censoring based on one of these models to construct an estimator for the joint concordance index. We prove that the proposed estimator is consistent. We introduce a variable importance ranking approach that we call stepwise competing risks regression. In this approach, we train a competing risks regression model and use backward elimination, i.e. drop the variable that leads to the least change in the joint concordance (See Figure \ref{figure1}). We also carry out experiments on real and synthetic datasets to establish the utility of the proposed metric.
	
	% and we show that the difference between the estimator and the joint concordance converges to a normal distribution. We show that the results that are presented for the joint concordance also extend to the generalized concordance. We  also propose  a variable importance ranking procedure based on the joint concordance that relies on backward elimination and stepwise regression. 

%	
%	 In the first set of experiments on synthetic datasets we evaluate the performance of the estimator in terms of the root mean squared errors, the standard errors, and the bias. 
%	We  show that a model selected based on the joint concordance has a   higher chance to correctly predict the event type and the event time for a subject in comparison to a model based on the existing metrics.  In addition, we show that the covariate ranking based on the  joint concordance for predicting  the CHD vs. STR  deaths ranks cholesterol, which is a CHD-specific risk factor and helps predict CHD deaths, to be the highest.  However, the ranking based on the standard approaches assign cholesterol a much lower rank.
%	
	%The code for the joint concordance index is available at \cite{kartik_code1}. 
	\subsection{Software}
	\label{software}
	
	The code for the joint concordance index is available at \footnote{\url{https://github.com/ahujak/Concordance_Based_Metrics_Survival}}. We also developed an application, which is  available at \footnote{\url{https://mlinterpreter.shinyapps.io/concordance/}}. The input/output for application is a) Input: Upload a standard competing risks dataset and select the model: Fine-Gray or Cause-Specific Cox model.  b)	Output: performance of the model- concordance for each cause, accuracy and the joint concordance. 
	
%	The above application can help the users directly select the model that seems more appropriate.
	
	\section{Existing Concordance Index for Competing Risks and its limitations}
	\label{existing_cindex}
	%Competing risks survival analysis  models have generated a wide amount of interest in the recent years [1][2]. These models are useful in various settings: i) Diseases such as different types of cardiac disease (for instance, ischemic heart disease and cerebrovascular heart disease) have both shared risk factors and separate risk factors [3].  Competing risks models enable the analysis of risk factors for the different outcomes using subdistribution functions, ii) In older populations, the development of treatment plans for cardiovascular disease should take into account the risks of mortality due to competing event types [4],  iii) In critical care settings,  many models have been developed to predict the length of stay and the status at discharge (dead or alive) [5][10][11]. Many standard models [10][11] treat the discharged alive information as independent censoring. However, it has been shown [5] that censoring in these cases is dependent on the covariates and not independent.  Competing risks models have been used in these settings to model the different types of events - death and discharge. 
	\subsection{Definition}
	In this section, we formally describe the most commonly used metric in SA-CR [6] for evaluating prognostic models. We begin by considering an uncensored dataset. We consider a dataset $\mathcal{D}$ comprising of the survival (event time) data for $n$ subjects who have been followed up for a finite amount of time. Let $\mathcal{D} = \{X_i, T_i, D_i,\;  i=1,..,n\}$, where $X_i \in \mathcal{X}$ is a $d$-dimensional vector of covariates associated with the subject $i$ (for instance, the information collected at baseline such as gender, age, etc.), $T_i \in \mathbb{R}^+$ is the time until an event occurred, and $D_i \in \mathcal{K}$ is the type of event that occurred. The set $\mathcal{K} = \{E_1,E_2\}$ is a finite set of  competing events that could occur to a subject $i$, where
	$E_1$ is the event of the first type and $E_2$ is the event of the second type.  The Cumulative Incidence Function (CIF) [7]  is the probability of occurrence of a particular event in $d\in \mathcal{K}$ by time $t$, and is given by 
	$F_{d}(t|X) = Pr(T\leq t, D=d|X)$, where $X\in \mathcal{X}$ is the covariate vector of the subject, $T$ is the event time, and $D$ is the event type.

	\subsubsection{Time-dependent event-specific concordance index}  The concordance index  measures a model's ability to discriminate the subjects.  The concordance index is defined for each event type separately.  
	Suppose a model $M$ predicts the risk of event $d$ until time $t$ to be $M(X,t,d)$.  Consider an independent test set of i.i.d. realizations of $(X_i , T_i , D_i )$ from the joint distribution of the covariates and the competing risks outcome.  For a random pair of subjects $(X_i , T_i , D_i )$ and $(X_j , T_j , D_j)$, the time dependent concordance for the event type $E_k$ is defined as
			\begin{equation}
	\begin{split}
	\mathcal{C}(t,k) \coloneqq Pr\Big(M(X_i,t,E_k) > M(X_j,t,E_k) \Big| \big[D_i=E_k \;\textit{and}\;T_i \leq t\; 
	\textit{and}\; (T_i<T_j \;\textit{or}\; D_j\not= E_k)\big] \Big) \\
	\end{split}
	\label{concordance}
	\end{equation}
	 For the pair of subjects $(X_i , T_i , D_i )$ and $(X_j , T_j , D_j)$ in \eqref{concordance}, the first subject $(X_i , T_i , D_i )$ would be in greater need of treatment for the event type $d$ than the second subject $(X_j , T_j , D_j)$  if they experienced the event of interest $(D_i =d)$ and the second subject experienced the event of interest later
	$(T_i < T_j \; \textit{and}\; D_j =d)$ or not at all $(D_j\not= d)$.
	The concordance vector is defined as a vector consisting of the time-dependent concordance for every event type and it is given as $\bar{\mathcal{C}}(t) = [\mathcal{C}(t,1), \mathcal{C}(t,2)]$. Note that the definition trivially extends to more than two event types.
	
	%\begin{figure*}
	%	\begin{equation}
	%	\begin{split}
	%	\mathcal{C}(t,k) = Pr\Big(M(X_i,t,E_k) > M(X_j,t,E_k) \Big|\big[D_i=E_k \;\textit{and}\;T_i \leq t\; 
	%	\textit{and}\; (T_i<T_j \;\textit{or}\; D_j\not= E_k)\big] \Big) \\
	%	\end{split}
	%	\label{concordance}
	%	\end{equation}
	%\end{figure*}

	\subsection{Limitations of the existing concordance index} \label{limitations_cindex} Each element of the concordance vector $\bar{\mathcal{C}}(t)$ defined above consists of information regarding a model's discrimination ability for each event type. However, it does not consist of information on whether the model is good at predicting the event type as well. In many applications, the evaluation of a model's ability to \textit{jointly} predict the event type and the event time is critical. For instance, treatment planning in multimorbid populations [8],  resource planning in critical care [5].

	\section{Joint Concordance Index}
	\label{jcindex}
	\subsection{Naive solution}
	We first describe an intuitive solution to overcome the limitations of concordance index discussed in Section \ref{limitations_cindex}. Define a model $M$'s prediction for the event type up to time $t$ for subject $X_i$ as $M_c(X_i,t)$. We define the accuracy of a model as  $\mathcal{A}(t) = Pr(M_c(X_i,t)=D_i | T_i \leq t)$

	In the definition, we condition on $T_i\leq t$ because we can evaluate a model's prediction only for the subjects who experienced the event before the stated time horizon, i.e., $t$.
	We construct a vector of  the concordance index for all the event types, and the accuracy  given as 
	$\mathcal{V}(t) = [\mathcal{C}(t,1), \mathcal{C}(t,2),  \mathcal{A}(t)]$
	
	Intuitively, it might seem that this vector is sufficient to capture a model's ability to make the  joint prediction of the event type and event time because the accuracy contains information about the ability of a model to predict the event type and the concordance captures the ability of the model to discriminate the event time for every event type separately. This solution is appealing because it is simple, but it has limitations that we  discuss next.   Suppose that a model makes a correct prediction of the event type for a subject $X_i$. Therefore, the condition inside the probability in $\mathcal{A}(t)$ is true for this subject. However, it is possible that for the same subject the discrimination condition inside \eqref{concordance} is not satisfied, which implies that the model is good at predicting the event type but not the event time for the predicted event type for this subject.  Therefore, concordance and accuracy (that comprise the vector $\mathcal{V}(t)$) evaluate the marginal probabilities and not the joint probabilities.   The joint evaluation is not trivial as the accuracy, and the concordance events are neither  independent nor  imply one another. 
	
	\subsection{Definition}
	\label{joint_concordance_sec}
	The joint concordance index is the probability that a given model  accurately predicts the event type for a subject while also ranking that subject's risk  correctly among the other subjects. 
	The expression for the joint concordance is given as 
		\begin{equation}
		\begin{split}
		& \mathcal{JC}(t) \coloneqq Pr\Big(M(X_i,t, D_i)>M(X_j,t,D_i)\; \& \; M_c(X_i,t) = D_i  \Big| \big[T_i \leq t \; \& \; \big(T_i<T_j \;\textit{or}\; D_i\not= D_j\big)\big]\Big) \label{joint_concordance}
		\end{split}
		\end{equation}

	In the equation \eqref{joint_concordance}, the model's ability to predict the event type for the subject and  discriminate that subject from the other subjects  is jointly evaluated.

	\subsection{Relationship with the existing metrics and interpretation}
	In this section, we study the relationship between the joint concordance and the existing metrics. We  decompose the joint concordance  into two terms that are easier to interpret as follows

		\begin{equation}
		\begin{split}
		\mathcal{JC}(t) = 	& \Big(    \overbrace{Pr\Big(M(X_i,t, D_i)>M(X_j,t,D_i) \Big| 
			D_i= M_c(X_i,t) \;\&\; T_i \leq t \; \& \; [T_i<T_j \;\textit{or}\; D_i\not= D_j]\Big)}^{\text{Concordance conditional on the correct prediction of the event type}}  \\ &\overbrace{Pr\big( M_c(X_i,t) =D_i \Big| T_i \leq t \; \& \; [T_i<T_j \;\textit{or}\; D_i\not= D_j]\big) \Big)}^{\text{Accuracy*}}\
		\end{split}
		\label{joint_concordance_decomposition}
		\end{equation}

	In equation \eqref{joint_concordance_decomposition}, the first term is concordance conditional on the event that the model correctly predicts the event type \eqref{concordance}. The conditional concordance \eqref{joint_concordance_decomposition} is evaluated for the subjects for which the events are predicted correctly unlike the standard concordance  \eqref{concordance} that is evaluated even for the subjects for which the wrong event type was predicted. The second term in the decomposition \eqref{joint_concordance_decomposition} is similar to the accuracy $\mathcal{A}(t)$. The difference between the accuracy term in \eqref{joint_concordance_decomposition} and $\mathcal{A}(t)$ is that the event in the conditional probabilities is different. From \eqref{joint_concordance_decomposition}, it might seem that the  joint concordance can be expressed as a function of the existing metrics- concordance and accuracy. However, this is not the case. In the next proposition, we show that the joint concordance cannot be expressed as a function of the existing metrics defined in $\mathcal{V}(t)$.

	%Let $\boldsymbol{w}$ be a weight vector, then $\boldsymbol{w}^t\mathcal{V}(t)  $ is a weighted sum of the existing metrics. 
	\begin{proposition}
		\textbf{Joint concordance vs. Existing metrics.}
		$\exists$ no function $f:\mathbb{R}^3 \rightarrow \mathbb{R}$ such that 
		$\mathcal{JC}(t) = f(\mathcal{V}(t)) $ 
	\end{proposition}
	The proofs to all the propositions are in the Appendix Section.

	\subsection{Estimators of joint concordance}

	\textbf{Weighted estimator to account for censoring.}  In the description of the dataset in Section  \ref{existing_cindex}, we assumed that there was no censoring.  We now introduce censoring variables. $C_i$ is defined as the censoring time for subject $i$. For subject $i$ we observe $X_i, \tilde{T}_i, \tilde{D}_i, \Delta_i$, where $\tilde{T}_i = \min\{T_i, C_i\}$ is the event time, $\Delta_i = I(T_i\leq C_i)$, type of event $\tilde{D}_i =\Delta_i D_i$.  We make the standard assumption that the censoring is independent of other variables conditional on the covariates. The probability that the subject $i$ is uncensored up to time $t$ is given as $G(t) = Pr(C_i>t | X_i)$.  We  use the inverse probability of  censoring weighted  (IPCW) (See [1]) to adjust for the bias that is introduced by censoring.    We can use different models to estimate the censoring bias; we denote the estimated model of censoring as $\hat{G}$. The two most natural choices for estimating the censoring models are:  i) Kaplan-Meier estimator  and ii) Cox model estimator of the censoring distribution. We provide some notation below that is needed to define the estimator.

	$\tilde{A}_{ij} = I(\tilde{T}_i < \tilde{T}_j)$, $\tilde{B}_{ij}(d) = I(\tilde{T}_i>\tilde{T}_j, \tilde{D}_j\not =d )$, $\tilde{N}_i(t,d) = I(\tilde{T}_i \leq t, \tilde{D}_i=d)$,  $\tilde{C}_{ij}(d) = I(\tilde{T}_i <\tilde{T}_j \;\text{or} \; \tilde{D}_j \not= d)$, $Q_{ij}(t,d) = I(M(X_i,t,d)>M(X_j,t,d)\; \&\; M_c(X_i)=d ) $  $\hat{W}_{ij}^{1} = \frac{1}{\hat{G}(T_i|X_i) \hat{G}(T_i |X_j)}$ $\hat{W}_{ij}^{2} = \frac{1}{\hat{G}(T_i|X_i) \hat{G}(T_j |X_j)}$

	The weighted estimator is defined as
	\begin{equation}
	\begin{split}
	\hat{\mathcal{JC}}_{wtd}(t) \coloneqq \frac{\sum_{d \in \mathcal{K}}\sum_{i,j}  (\tilde{A}_{ij}\hat{W}_{ij}^{1} + \tilde{B}_{ij}(d)\hat{W}_{ij}^{2})N_{i}(t,d) Q_{ij}(t,d)}{\sum_{d \in \mathcal{K}}\sum_{i,j} (\tilde{A}_{ij}\hat{W}_{ij}^{1}+ \tilde{B}_{ij}(d)\hat{W}_{ij}^{2})N_{i}(t,d) }
	\label{weighted_estimator}
	\end{split}
	\end{equation} 
	The denominator in \eqref{weighted_estimator} counts the number of pairs in the dataset that satisfy the definition of the event in the conditional probability in \eqref{joint_concordance}. The numerator in \eqref{weighted_estimator} counts the number of jointly concordant pairs (pairs for which the prediction of the model is accurate and the concordance condition holds).

	%\end{figure}
	In the next proposition, we show that the weighted estimator \eqref{weighted_estimator} is consistent. For the next proposition, we require that the model for censoring is correctly specified. The same assumptions were also made in [6].
	
	\begin{proposition}  \textbf{Properties of the estimator}
		$\hat{\mathcal{JC}}_{wtd}(t)$ is a consistent estimator of the joint concordance $\mathcal{JC}(t)$.

	\end{proposition}
	\subsection{Variable importance ranking}
	\label{vimp} First we highlight the limitations of the existing approaches for ranking the variables with respect to the overall risk profile and propose an alternate approach based on joint concordance that overcomes these limitations. 
	
	\subsubsection{Existing approaches} We describe the two most common approaches that are used for variable importance ranking-  standardized regression coefficients based approaches [9] and the stepwise regression based approaches [10].  The existing works [11] [12] rank the covariates for the overall risk profile by lumping the different event types into one common group and then using the standardized regression coefficients or the stepwise regression methods to rank the covariates with respect to the risk of the lumped event (See Figure \ref{figure1}). In the comparisons to follow, for the stepwise regression methods we use concordance index defined in  \eqref{concordance} as the measure that is compared in each step. We also contrast our results with the standardized regression coefficient based approach.

	\begin{figure}
		\begin{center}
			\includegraphics[trim= 0mm 45mm 50mm 0mm, width=4 in]{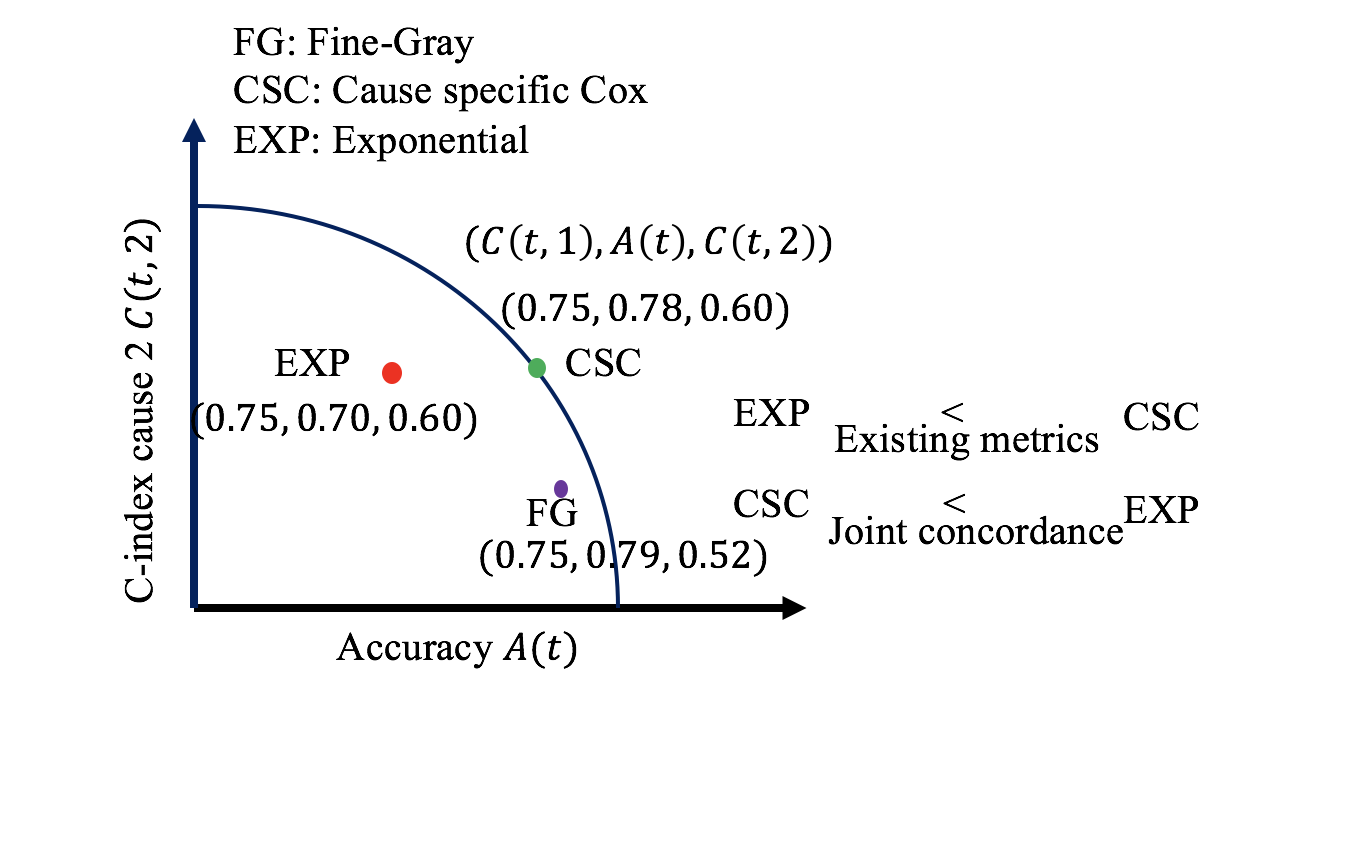}
			
			\caption{Comparing different models in terms of existing
				metrics for the synthetic setting}
		\end{center}
		\label{Comparison_joint_existing}
	\end{figure}
	
	\subsubsection{Stepwise competing risks (CR) regression approach} We refer to the proposed approach as stepwise competing risks (CR) regression approach. First  we first train a competing risks model on all the variables.  We use  backward elimination with stepwise regression  with joint concordance as the metric. In each step of the backward elimination, we compute the joint concordance for the trained model. We drop the variable that leads to the least amount of change in the joint concordance when dropped (See Figure \ref{figure1}). The same procedure is repeated after dropping the variable. Note that the least important variable is dropped first and the most important variable is dropped last.
	
	\section{Experiments}
	\label{experiments_section}
	In this section, we first discuss the synthetic data experiments and then discuss the real data experiments. These experiments are carried out with three goals in mind: \begin{enumerate}\item Existing metrics are not sufficient for joint evaluation
		
		\item Existing variable importance ranking often fail to rank the factors with respect to the overall risk 
		\item Compute the efficiency of the weighted estimator
		
	\end{enumerate}
	All the experiments were conducted in the R.

	\subsection{Synthetic Data Experiments}
	\label{syn_dat_exp}
	
	\textbf{Synthetic  experiment setting.} We use an experiment setting that is very similar to [6]. The covariate of a subject is $X\in \mathbb{R}$. It is drawn from a standard normal distribution.  Suppose that there are three event types - event of type 1,  event of type 2, and censoring. We use an accelerated failure time model [7] to model the event time. The latent time for event type $k$ is  $T_k$ and it is drawn from an exponential distribution with arrival rate $\lambda_k(t|X) = \lambda_k(t) \text{exp}(\beta_k X)$ for $k\in \{0,1\}$ where the event type $k=0$ is censoring and event type $k=1$ is the event of  type 1.  The latent time for the event of second type is $T_2$ and it is also drawn from an exponential distribution with parameters $\lambda_2(t|X) = \lambda_2(t) \text{exp}(\beta_2  \text{cos}(X))$
	The observed event time is $T = \min_{k\in \{0,1,2\}}\{T_k\}$ and the observed type of event is $d = \arg\min_{k\in \{0,1,2\}}\{T_k\}$. The parameters above are chosen as follows $\lambda_{1}(t) = 1, \lambda_{2}(t) =2 , \beta_1 = 1, \beta_2 =1$. For $\beta_0$, we set two different values, $\beta_0 = 0$ for the covariate independent censoring  and  $\beta_0 =1$ for the  covariate dependent censoring. We set $\lambda_0(t)$  such that the proportion of the right censored data points is $50\%$ and $75 \%$. We compare the different models in terms of the existing metrics and the joint concordance at the $75\%$ quantile of the  times. We use three models for comparisons here: i) Cause-specific Cox model (CSC) \footnote{We used the  riskRegression package in R for the CSC model.}  ii) Fine-Gray model (FG)\footnote{We used the cmprsk package in R for the FG model.}  [13], and  iii) the exponential model (EXP) ($M(X,t,1) = \text{exp}(X)$, $M(X,t,2)  =2\text{exp}(-\text{abs}(X))$, where $\text{abs}(X)$ is the absolute value of $X$).  
	
	\textbf{Model comparisons} Our goal is to show that the existing metrics can lead to the selection of models that are bad for joint prediction of the event type and event time.  First, we compute the exact  values for all the metrics (concordance, accuracy, and joint concordance) using a large data set of  $100,000$ subjects for the synthetic experiment setup described above but in the absence of censoring. We compare the models in terms of standard metrics (concordance index for each event type [6] and the accuracy) $\mathcal{V}(t)$. We focus on the comparison between the CSC model and the EXP model. Based on the  standard metrics (in Figure 2 and  Table \ref{model_comparison_synthetic_table}), the CSC model  seems to be better than the EXP model. However, when we compare the joint concordance,  we find that the EXP model is better even though the CSC model Pareto dominates the EXP model in terms of existing metrics.   EXP model has a 4 $\%$ higher chance of correctly predicting both the event type and event time for a subject. We use the decomposition in \eqref{joint_concordance_decomposition} to get further insights into this comparison. The concordance conditional on accuracy for the EXP model is 0.74, and the accuracy is 0.70. The concordance conditional on accuracy for CSC model is 0.61, and the accuracy is 0.78. Although the CSC model can predict the event type in more cases, it is very poor in discriminating the subject for which it predicts the event correctly from other subjects. Poor discrimination implies that the event time predictions are also poor.  Therefore, the CSC  model is worse in comparison to the EXP model for the joint prediction of the event type and event time. Hence, the existing metrics can lead to poor model selection.

	\textbf{Efficiency of the weighted estimators:}  In this section, we evaluate the efficiency of  the weighted estimators. We use the synthetic experiment setting described above that was also used to compare the models. We use the Kaplan Meier estimator for estimating the censoring distribution.  We use the simulated datasets of size $1000$ and $5000$. We compute the RMSE, SE, and the bias by averaging over 100 such datasets and compare them in Table \ref{tablese}. All the comparisons that are carried out are in-sample. In Table \ref{tablese}, we see that  when the censoring rates are lower, the weighted estimator has a lower RMSE and  bias.
	%The comparisons when the censoring is covariate dependent  are  in the Appendix H.
	%
	%
	%\begin{table}
	%	\caption{Root mean squared error (RMSE), Standard Error (SE) and Bias for  the estimators: Covariate independent censoring}
	%	\centering
	%	\begin{tabular}{ | p{0.9cm} |p{1cm}| p{1.4cm} | p{1.2cm}| p{1cm}| p{0.9cm}| p{1.2cm}| }	
	%		\hline
	%		\textbf{Model} & $\boldsymbol{\mathcal{JC}(t)}$ & \textbf{Censoring} & \textbf{No. of samples}  &  \textbf{RMSE}  & \textbf{SE}  & \textbf{Bias} \\ \hline
	%		
	%		CSC & $0.48$ & 50 \%  & $1000$ &  0.0249 & 0.0247 & 0.0039\\ 
	%		FG & $0.46$  & 50 $\%$ & $1000$  & 0.0338 & 0.0265 & -0.0211\\ 
	%		EXP & $0.52$    & 50 $\%$ & $1000$ &  0.0179 & 0.0160 & 0.0081  \\ 
	%		
	%		CSC & $0.48$  & 50 \%  & $5000$ &  0.0177 & 0.0137 & 0.0113\\ 
	%		FG & $0.46$ & 50 $\%$ & $5000$  & 0.0259 & 0.0106 & -0.0236 \\ 
	%		EXP & $0.52$    & 50 $\%$ & $5000$ & 0.0103 & 0.0067 &  0.0082  \\ \hline 
	%
	%	\end{tabular}
	%	\label{table_se}
	%\end{table}
	\begin{table}	
		\centering
		\caption{RMSE, SE and BIAS for the estimator}
		\begin{tabular}{ | p{1cm} |p{1cm}| p{1.3cm} | p{1.8cm}| p{1cm}| p{1cm}| p{1cm}| }	
			\hline
			\textbf{Model} & $\boldsymbol{\mathcal{JC}(t)}$ & \textbf{Censoring} & \textbf{$\#$ of samples}  &  \textbf{RMSE}  & \textbf{SE}  & \textbf{Bias} \\ \hline
			
			CSC & $0.48$ & 50 \%  & $1000$ &  0.0249 & 0.0247 & 0.0039\\ 
			FG & $0.46$  & 50 $\%$ & $1000$  & 0.0338 & 0.0265 & -0.0211\\ 
			EXP & $0.52$    & 50 $\%$ & $1000$ &  0.0179 & 0.0160 & 0.0081  \\ 
			
			CSC & $0.48$  & 50 \%  & $5000$ &  0.0177 & 0.0137 & 0.0113\\ 
			FG & $0.46$ & 50 $\%$ & $5000$  & 0.0259 & 0.0106 & -0.0236 \\ 
			EXP & $0.52$    & 50 $\%$ & $5000$ & 0.0103 & 0.0067 &  0.0082  \\ 
			
			CSC & $0.48$ & 75 \%  & $1000$ & 0.0350 & 0.0338 & -0.0095\\ 
			FG & $0.46$  & 75 $\%$ & $1000$  &   0.0374 & 0.0319 & -0.0192\\ 
			EXP & $0.52$    & 75 $\%$ & $1000$ &  0.0308 & 0.0231 & 0.0205  \\ 
			
			CSC & $0.48$  & 75 \%  & $5000$ &  0.0231 & 0.0231 & 0.0020 \\ 
			FG & $0.46$  & 75 $\%$ & $5000$  &  0.0326 & 0.0189 & -0.0266 \\ 
			EXP & $0.52$    & 75 $\%$ & $5000$ & 0.0202 & 0.0089 & 0.0180\\ \hline
		\end{tabular}
		\label{tablese}
	\end{table}	
	
	%  Note that when the censoring is covariate dependent both naive and weighted estimators are not consistent.   
	% In Table 1, we can see  that in general when the censoring is covariate dependent, naive estimator is better in general. 
	%\begin{figure}
	%	\begin{center}
	%		\includegraphics[trim=0mm 0mm 0mm 0mm, width=3.2in]{Comparison_new_dec18.png}
	%		\caption{Comparing different models in terms of existing metrics for the synthetic setting}
	%		\label{Comparison_joint_existing}
	%	\end{center}
	%\end{figure}
	
	\begin{table}
		\renewcommand{\arraystretch}{1.2}
		\caption{Model comparisons in terms of existing metrics vs joint concordance: synthetic data}
		\centering
		\begin{tabular}{ | p{1cm} | p{1cm} | p{1cm}| p{1cm}| p{1cm}|}	
			\hline
			\textbf{Model}  &  $\mathcal{C}(t,1)$  & $\mathcal{C}(t,2)$ & $\mathcal{A}(t)$ &  $\mathcal{JC}(t)$ \\ \hline
			FG   & $0.75$ & $0.52$ & $0.79$ & $0.46$ \\ 
			CSC & $0.75$ & $0.60$ & $0.78$    & $0.48$  \\ 
			EXP & $0.75$ & $0.60$ & $0.70$     & $0.52$  \\ \hline
		\end{tabular}
		\label{model_comparison_synthetic_table}
	\end{table}
	
	\begin{table}
		\renewcommand{\arraystretch}{1.2}
		\caption{Model comparisons in terms of existing metrics vs joint concordance: CHD deaths vs HDUE deaths}
		\centering
		\begin{tabular}{ | p{1cm} | p{1.4cm} | p{1.4cm}|p{1cm} |p{1cm}|}	
			\hline
			\textbf{Model}  &  $\mathcal{C}(t,CHD)$  & $\mathcal{C}(t,STR)$ &  $\mathcal{A}(t)$ & $\mathcal{JC}(t)$ \\ \hline
			FG   & $0.76$ & $0.79$ & $0.56$ & $0.45$ \\ 
			CSC & $0.76$ & $0.79$ & $0.55$  & $0.43$  \\ \hline
		\end{tabular}
		\label{model_comparison_real_table}
	\end{table}
	
	\begin{table}
		\renewcommand{\arraystretch}{1.2}
		\caption{Model comparisons in terms of existing metrics vs  proposed metric: HDUE deaths vs STR deaths}
		\centering
		\begin{tabular}{ | p{1cm} | p{1.8cm} | p{1.5cm}| p{1cm}| p{1cm}| }	
			\hline
			\textbf{Model}  &  $\mathcal{C}(t,HDUE)$  & $\mathcal{C}(t,STR)$  &  $\mathcal{A}(t)$ & $\mathcal{JC}(t)$ \\ \hline
			FG   & $0.73$ & $0.79$ & $0.68$ & $0.55$ \\ 
			CSC & $0.76$ & $0.79$ & $0.68$ & $0.55$  \\ \hline
		\end{tabular}
		\label{model_comparison_real_table2}
	\end{table}

	\subsection{Real Data Experiments} 
	
	\label{model_comp_real_data}
	In this section, we use  real datasets to illustrate real use cases of the joint concordance index. We use cardiovascular risk dataset and SEER dataset.

	% is better at jointly predicting the type of event and event time for 4 $\%$  more subjects than the CSC model even though the CSC model does a better job at predicting the event type only in comparison to the EXP model. 
	
	%In addition to comparing with the event type-specific concordances and the accuracy we also compared the concordance when all the event types are lumped into a single event type (see Table \ref{model_comparison_synthetic_table} $\mathcal{C}^{*}(t)$). We find that in terms of this metric the models are indistinguishable.

		\textbf{Cardiovasuclar Risks Dataset} The dataset [11] comprises of 1712 men (aged 40-59 years in 1960) from Italian Rural
	Areas of the Seven Countries Study [11]. During the 50-year follow-up, there were 12 different causes
	of death recorded: 318 due to Coronary Heart Disease (CHD), 162 to Heart Disease of Uncertain
	Etiology (HDUE), 225 to Stroke (STR) and another 964 due to miscellaneous causes. Covariates
	measured at baseline were: Age, Arm Circumference, Cigarettes, Body Mass Index, Diabetes,
	Corneal Arcus, Serum Cholesterol,  Blood Pressure, Heart Rate and Vital Capacity.
	
	\textbf{SEER Datasets}
	We extracted 2 cohorts from the Surveillance, Epidemiology, and End Results
	(SEER) cancer registries, which cover approximately 28\%
	of the US population [14]. In the first cohort the outcome is the time to cancer deaths versus the time to non-cancer deaths. In the second cohort, the outcome is time to digestive cancer deaths versus the time to breast cancer deaths.  There are 81 covariates and here we specify some important of them: Age, Age at Diagnosis, Histology, Tumor size, Family history, Tumor grade, Race,  etc. 
	
%	\textbf{Model Comparisons} In this section, we compare the ranking of models obtained using existing metrics as opposed to the ranking based on the joint concordance. 

	%We compare the predictions of the models at the time horizon of 15 years. 

	\textbf{Model comparisons on cardovascular risks dataset:} We estimate two models, the CSC model and the FG model.   
	In the first comparison, the two competing events are the death due to CHD and the death due to STR.   The comparisons in Table \ref{model_comparison_real_table}    reveal that the FG  model and the CSC model are similar in terms of the standard concordance metrics and there is a small difference in terms of the accuracy. In this comparison, the FG model Pareto dominates the CSC model in terms of the existing metrics. We find that the FG model is also better in terms of the joint concordance. 
	
	For the second comparison given in Table \ref{model_comparison_real_table2}, the two competing events are the death due to HDUE and the death due to STR.  The comparisons in Table \ref{model_comparison_real_table2}  reveal that the CSC model Pareto dominates the FG model. However, there is no difference (statistically significant) between the joint concordance of both the models.

	The takeaway from the comparisons in Tables \ref{model_comparison_real_table}, \ref{model_comparison_real_table2} is that a comparison in terms of the existing metrics is not sufficient to deduce the performance in terms of the joint concordance. Moreover, selecting a model based on the existing metrics can lead to selection of a model that performs worse in terms of joint prediction of event type and event time.

	\textbf{Model comparisons on SEER datasets:} We estimate two models, the CSC model and the FG model.   
	For the third comparison given in Table \ref{model_comparison_real_table3}, the two competing events are  death due to cancer and death due to any other cause. The comparisons in Table \ref{model_comparison_real_table3} reveal that the CSC model and the FG model are similar in terms of the existing metrics. The CSC model performs better in terms of the joint concordance. 
	
	For the fourth comparison given in Table \ref{model_comparison_real_table4}, the two competing events are death due to breast cancer and death due to digestive cancer. In terms of the existing metrics, the CSC model seems to perform better. In terms of the joint concordance, the FG model is better.  
	From Table \ref{model_comparison_real_table3}, Table \ref{model_comparison_real_table4} we can conclude that if one selected a model just based on the existing metrics then it can lead to selection of a poor model in terms of joint predictions.
	\begin{table}
		\renewcommand{\arraystretch}{1.2}
		\caption{Model comparisons in terms of existing metrics vs joint concordance: Cancer deaths vs Non-Cancer deaths}
		\centering
		\begin{tabular}{ | p{1cm} | p{1.4cm} | p{1.6cm}| p{1.4cm}| p{1cm}|}	
			\hline
			\textbf{Model}  &  $\mathcal{C}(t,CAN)$  & $\mathcal{C}(t,NCAN)$ & $\mathcal{A}(t)$  & $\mathcal{JC}(t)$ \\ \hline
			FG   & $0.71$ & $0.61$ & $0.70$ &  $0.47$ \\
			CSC & $0.71$ & $0.61$ & $0.70$  &  $0.50$  \\ \hline
		\end{tabular}
		\label{model_comparison_real_table3}
	\end{table}
	
	\begin{table}
		\renewcommand{\arraystretch}{1.2}
		\caption{Model comparisons in terms of existing metrics vs joint concordance: Breast Cancer vs Digestive Cancer deaths}
		\centering
		\begin{tabular}{ | p{1cm} | p{1.6cm} | p{1.6cm}| p{1.4cm}| p{1cm}|}	
			\hline
			\textbf{Model}  &  $\mathcal{C}(t,BCAN)$  & $\mathcal{C}(t,DCAN)$ & $\mathcal{A}(t)$  & $\mathcal{JC}(t)$ \\ \hline
			FG   & $0.82$ & $0.87$ & $0.72$ &  $0.82$ \\ 
			CSC & $0.83$ & $0.87$ & $0.93$  &  $0.78$  \\ \hline
		\end{tabular}
		\label{model_comparison_real_table4}
	\end{table} 
	%We want to point out that this does not mean that the existing metrics are not good. For instance, if the goal of the study is to focus only on discrimination of subjects with respect to a certain event type, then the existing metrics are sufficient.
	%In the above comparisons, we did not learn joint models for all the twelve event types in the dataset because the dataset was too small to permit learning an accurate joint model. 

	\textbf{Variable Importance Ranking} In this section, our goal is to compare the standard approaches for variable importance ranking with the proposed stepwise CR regression (already described in the Section \ref{vimp}).  We used the same real dataset that we described in Section \ref{model_comp_real_data}. We carry out two comparisons: CHD deaths vs. STR deaths and HDUE deaths vs. STR deaths. We use the FG model to rank the risk factors. 
	
	In the first comparison given in Table \ref{variable_importance_table},  we compare the risk factor rankings when the two events are the CHD deaths and the STR deaths. 
	We show that the ranking arrived at using the stepwise CR regression can be very different than the ranking arrived at using the standard approach based on the stepwise regression.  We see that the proposed approach ranks cholesterol to be the highest, unlike the standard approach (cholesterol is ranked at seventh). Cholesterol is a strong event-specific risk-factor; it matters much more for the  CHD deaths in comparison to the STR deaths (this is well known in the clinical literature [11][12]).  The standard approach can miss such important risk factors. We also ranked the variables using  the standardized regression based approach and we obtained the same conclusions.

	In the second comparison in Table \ref{variable_importance_table1}, we compare the risk factor rankings when the two events are the HDUE deaths and the STR deaths. We show that the ranking arrived at using the joint concordance index is not very different in comparison to the ranking arrived at using the standard approach. This suggests that in this case for both the outcomes (HDUE deaths and STR deaths) the dataset does not contain risk factors that are only specific to one of the events.
	
	Therefore, from Tables \ref{variable_importance_table}, \ref{variable_importance_table1}, we can see that in the cases when the  dataset consists of risk factors that are exclusively specific to some events, the existing approaches can often fail to recognize their importance. On the other hand, the proposed approach is good at identifying the importance of these factors.
	
	\begin{table}
		\renewcommand{\arraystretch}{1.2}
		\caption{Compare variable importance: CHD  vs. STR deaths}
		\centering
		\begin{tabular}{ | p{1cm} | p{4cm} | p{4cm}|}	
			\hline  
			\textbf{Rank}  &  \textbf{Stepwise regression} & \textbf{Stepwise CR regression}\\ \hline
			1        & Blood Pressure  & Cholesterol \\ 
			2        & Age             & Blood pressure \\ 
			3        & Vital capacity  & BMI             \\ 
			4        & Arm circumference   &    Age   \\ 
			5        &  Corneal arcus      &   Diabetes \\ \hline
		\end{tabular}
		\label{variable_importance_table}
		
	\end{table}

	\begin{table}
			\renewcommand{\arraystretch}{1.2}
		\caption{Compare variable importance: HDUE  vs. STR deaths}
		\centering
		\begin{tabular}{ | p{1cm} | p{4cm} | p{4cm}|}	
			\hline  
			\textbf{Rank} &  \textbf{Stepwise regression} & \textbf{Stepwise CR regression}\\ \hline
			1        & Age & Blood Pressure \\ 
			2        & Vital capacity             & Age \\ 
			3        & Blood Pressure  & Vital capacity             \\ 
			4        & Arm circumference   &   Arm Circumference   \\ 
			5        &  Heart Rate     &   Heart Rate\\ \hline
		\end{tabular}
		\label{variable_importance_table1}
		
	\end{table}

	%
	% but do not provide a measure of separation for the overall risk profile. Measure D estimates the separation in the predicted log hazard ratios and it does not measure the model's ability to discriminate the individuals  

	%\subsection{Reclassification metrics} Net reclassification improvement \cite{leening2014net} \cite{wolbers2009prognostic} is a metric that has been used for understanding the added value of using a covariate in improving the prognostic accuracy of a model.  These reclassification metrics can also be used for variable importance ranking. In this work, we only extended 
	%
	%It would be useful to extend these reclassification measures to settings where the overall risk prediction of the subject is of interest.
	
	%
	\section{Conclusion}
	\label{conclusion_section}
	In SA-CR, existing metrics such as concordance and accuracy  do not evaluate a model based on its joint prediction of the event type and event time. We have  proposed a new metric that we call the joint concordance that overcomes the limitations of the existing metrics.   We have proposed an estimator for the joint concordance that adjusts for the bias that occurs due to censoring and we prove that it is consistent.  We have shown that the existing methods for variable importance ranking can often fail to recognize the importance of the event-specific risk factors, which are crucial for predicting the event type. We have introduced a new ranking method based on joint concordance that overcomes these limitations, which we call stepwise competing risks regression. 

	\section{Acknowledgement}
	
	We are grateful to Prof. Gary Collins (University of Oxford), Dr. Angela Wood (University of Cambridge) and Ahmed Alaa for their comments that helped improve this work.
	
	\section*{Appendix}
	
	\subsection*{Proof of Proposition 1} Consider the case when there are two event types. We  construct two models and compute their joint concordance and existing metrics. 
	
	The first model, i.e., Model 1, assigns the risk for a subject uniformly at random.  $R_{i}^{1}$ and $R_{i}^{2}$ follow a uniform distribution over $[0,1]$, where $R_{i}^{j}$ is the risk for individual $i$ for event type $j$.  The risks for all the indivduals are independent and identically distributed (i.i.d). 
	
	The second model, i.e., Model 2, assigns the risk $R_i^{1}$ to subject $i$ for event type $1$ from a uniform distribution over $[0,1]$. Model 2 assigns risk $R_{i}^{2} = 1-R_{i}^{1}$ for the second event.
	
	We compute concordance index of the  Model 1 and  the Model 2 for both event types as follows. 
%	We use the definition of the joint concordance from the main manuscript  to compute the concordance for the event type $k$  for Model 1.  
We need to compute	
	$Pr(R_{i}^{k}>R_{j}^{k}|(T_{i}\leq t ) \;\textit{and}\; \big(T_{i}<T_{j} \;\textit{or}\; D_{j}\not=k \big ))$.  $R_{i}^{k}-R_{j}^{k} $ is independent of $(T_{i}\leq t ) \;\textit{and}\; \big(T_{i}<T_{j} \;\textit{or}\; D_{j}\not=k \big )$). Therefore,  
	\begin{equation}
	\begin{split}
	&Pr(R_{i}^{k}>R_{j}^{k}|(T_{i}\leq t ) \;\textit{and}\; \big(T_{i}<T_{j} \;\textit{or}\; D_{j}\not=k \big )=  Pr(R_{i}^{k}>R_{j}^{k}) = \frac{1}{2}
	\end{split}
	\end{equation}
	
	 A similar simplifcation also leads to the same values for the concordances for the Model 2.

	We compute the accuracy for the Model 1. The accuacy for the event type $1$  for Model 1 is given  as follows
	\begin{equation}
	\begin{split}
	& Pr(R_{i}^{1}>R_{i}^{2}, D_i=1| T_i \leq t ) =  Pr(R_{i}^{1}>R_{i}^{2}, D_i=1) = \frac{1}{2}Pr(D_i=1|T_i\leq t) 
	\label{eqn1}
	 \end{split}
	 \end{equation}
	 
	 Similarly, we compute the accuracy for event type $2$ for Model 1 given as 
	 
	 \begin{equation}
	 \begin{split}
	 &Pr(R_{i}^{2}>R_{i}^{1}, D_i=2| T_i \leq t ) = Pr(R_{i}^{2}>R_{i}^{1}, D_i=2) = \frac{1}{2}Pr(D_i=2|T_i\leq t)
	 \end{split}
	 \label{eqn2}
	 \end{equation}
	 
	 If we add the above two equations \eqref{eqn1} and \eqref{eqn2}, we observe that the accuracy of the Model 1 is $\frac{1}{2}$.
	
	We now compute the accuracy for Model 2.   We use the definition of accuracy from the main manuscript to compute the accuracy for the event type $1$  for Model 2 as follows 
	\begin{equation}
	\begin{split}
&	Pr(R_{i}^{1}>1-R_{i}^{1}, D_i=1| T_i \leq t ) = Pr(R_{i}^{1}>1-R_{i}^{1}, D_i=1) = \frac{1}{2}Pr(D_i=1|T_i\leq t)
	\end{split}
	\end{equation}
	
	Similarly, we can compute the accuracy for event type $2$ for Model 2 and follow the same steps as that for Model 1 and find the accuracy to be equal to $\frac{1}{2}$.	We now compute the joint concordance for Model 2.  Suppose Subject $i$ experiences event type $1$. Let us assume that the model correctly predicted the event type, i.e.,  $R_i^1>\frac{1}{2}$.  Suppose Subject $j$ experiences the event $1$ at a later date or not at all. For Subject $j$, the value of the risk for event $R_j^1$ is also a uniform random variable. We need to compute the probability $Pr(R_{j}^{1}<R_{i}^{1}| R_{i}^{1}>\frac{1}{2})$. This probability can be written as the sum of two probabilities given as $Pr(R_{j}^{1}<\frac{1}{2}| R_{i}^{1}>\frac{1}{2}) = \frac{1}{2}$ and $Pr(R_{j}^{1}>\frac{1}{2}\; \& \; R_i^1<R_j^1| R_{i}^{1}>\frac{1}{2}) $ given as
	\begin{equation}
	\begin{split}
	& Pr(R_{j}^{1}>\frac{1}{2}\; \& \; R_i^1<R_j^1| R_{i}^{1}>\frac{1}{2}) = Pr(R_{j}^{1}>\frac{1}{2}, R_{j}^{1}>R_{i}^{1}| R_{i}^{1}>\frac{1}{2}) =\frac{1}{4}  
	\end{split}
	\label{jc_comp}
	\end{equation}
	Hence, the joint probability that the event type is predicted correctly and concordance condition also holds is\begin{equation}
	Pr(R_i^1>\frac{1}{2}, R_j^1<R_i^1)=\frac{1}{2}(\frac{1}{2}+ \frac{1}{4})  =\frac{3}{8}\end{equation}
	Therefore, the joint concordance for Model 2 is $\frac{3}{8}$.
	We now compute the joint concordance for Model 1. $$Pr(R_i^1>R_j^1\; \& \;R_i^1>R_i^2) =\frac{1}{3}$$ $$Pr(R_i^2>R_j^2\; \& \;R_i^2>R_i^1) =\frac{1}{3}$$
	The above is true because $R_i^1, R_i^2, R_j^1$ are i.i.d. These two conditions together with the fact that $R_i^{k}$ are uniform random variables that are i.i.d. lead to the fact that joint concordance for Model 1 is $1/3$.
	Suppose that the claim in the Proposition is false, i.e. there exists a function $f$ such that $\mathcal{JC}(t) = f(\mathcal{V}(t))$. In the above examples, Model 1 and Model 2 have the same values for $\mathcal{V}(t)$ equals to $[\frac{1}{2}, \frac{1}{2}, \frac{1}{2}]$. If the above identitiy is true, then the joint concordance for Model 1 and Model 2 is $f([\frac{1}{2}, \frac{1}{2}, \frac{1}{2}])$. But based on the computations above, Model 1 and Model 2 have different values for the joint concordance \eqref{jc_comp}. This leads to a contradiction. Therefore, there exists no such function. \hfill QED
	
	\textbf{Proof of Proposition 2:} 
	We want to simplify 
	\begin{equation}Pr(D_i =d \; \& \;T_i \leq t \; \& \; (T_i<T_j \;\textit{or}\; D_i\not= D_j)
	\label{eq1_prop2} \end{equation}
	
	We use 
	 $I(s<T_j \;\textit{or}\; D_j\in{d}^{c})  = 1- I(T_j\leq s \; \textit{and}\;D_j =d)$
	in \eqref{eq1_prop2} to obtain
	
	\begin{equation}
	\begin{split}
	&	Pr(D_i =d \; \& \;T_i \leq t \; \& \; (T_i<T_j \;\textit{or}\; D_i\not= D_j) |X_i, X_j)  =   \sum_{d}\int_{0}^{t}E_{X_i, X_j}\Big[(1-F_d(s|X_j))dF_{d}(s|X_i)\Big]
	\end{split}
	\label{simplify_JC_0}
	\end{equation}
	In the above \eqref{simplify_JC_0}, $F_d$ is the CIF  defined in Section \ref{jcindex}.

	We use this equation \eqref{simplify_JC_0} to simplify the expression for joint concordance as follows
		\begin{equation}
		\small
	\begin{split}
	&\mathcal{JC}(t) = Pr\Big(M(X_i,t, D_i)>M(X_j,t,D_i)\; \& \; M_c(X_i,t) = D_i \Big|T_i \leq t \; \& \; (T_i<T_j \;\textit{or}\; D_i\not= D_j)\Big)  \\ 
	%		&= \sum_{d}\frac{Pr\Big(M(X_i,t, d)>M(X_j,t,d)\; \& \; M_c(X_i,t) = d \; \&\; D_i=d\; \& \;T_i \leq t \; \& \; (T_i<T_j \;\textit{or}\; D_i\not= D_j)\Big)}{Pr(T_i \leq t \; \& \; (T_i<T_j \;\textit{or}\; D_i\not= D_j)) } \\
	& \sum_{d}\frac{E_{X_i, X_j}\Bigg[Pr\Big(M(X_i,t, d)>M(X_j,t,d)\; \& \; M_c(X_i,t) = d \; \&\; D_i=d\; \& \;T_i \leq t \; \& \; (T_i<T_j \;\textit{or}\; D_i\not= D_j)\Big| X_i,X_j\Big)\Bigg]}{Pr\Big(T_i \leq t \; \& \; (T_i<T_j \;\textit{or}\; D_i\not= D_j)\Big) }\\
	%		&	\text{Substituting expressions from the equation \eqref{simplify_JC_0} to simplify the above expression we get}\\
	&	\mathcal{JC}(t) = 	 \sum_{d}\frac{E_{X_i,X_j}\Bigg[I(M(X_i,t, d)>M(X_j,t,d)\; \& \; M_c(X_i,t) = d )\int_{0}^{t}(1-F_d(s|X_j))dF_{d}(s|X_i)\Bigg]}{\sum_{d}E_{X_i, X_j}\Bigg[\int_{0}^{t}(1-F_d(s|X_j))dF_{d}(s|X_i)\Bigg] }                
	\end{split}
	\label{simplify_JC1}
	\end{equation}

	 $\hat{G}$ is a consistent estimator for $G$ (follows from the assumption that the censoring model is correctly specified).  The weights $	\hat{W}_{ij}^{1} $ and $	\hat{W}_{ij}^{2}$  converge in probability to

	%	 From Slutsky's lemma that
	\begin{equation}
	W_{ij}^{1} = \frac{1}{G(\tilde{T}_i-|X_i)G(\tilde{T}_i|X_j)} 
	\label{wij1}
	\end{equation}
	
	\begin{equation}
	W_{ij}^{2} = \frac{1}{G(\tilde{T}_i-|X_i)G(\tilde{T}_i|X_j)}
	\end{equation}

	From law of large numbers and Slutsky's lemma we can conclude that  the estimator in \eqref{weighted_estimator} converges to \begin{equation}
	\small
	\begin{split}
	\hat{\mathcal{JC}}_{wtd}(t) &=	 \sum_{d}\frac{\sum_{i}\sum_{j}\Big[A_{ij}\hat{W}_{ij}^{1} + B_{ij}(d)\hat{W}_{ij}^{2}\Big]N_{i}(t,d)I\Big(M(X_i,t, d)>M(X_j,t,d)\; \& \; M_c(X_i,t) = d \Big)}{\sum_{d}\sum_{i}\sum_{j}\Big[A_{ij}\hat{W}_{ij}^{2} + B_{ij}(d)\hat{W}_{ij}^{2}\Big]N_{i}(t,d) } \\ 
	\;\;\;\;\;\;\;\;\;\;\;\;\;\; &	\rightarrow	 \sum_{d}\frac{E_{X_i, X_j}\Bigg[E\Big[\big[A_{ij}W_{ij}^{1}N_{i}(t,d) + B_{ij}(d)W_{ij}^{2}N_{i}(t,d)\big]\Big| X_i, X_j\Big]I\Big(M(X_i,t, d)>M(X_j,t,d)\; \& \; M_c(X_i,t) = d \Big)\Bigg]}{\sum_{d}E_{X_i, X_j}E\Bigg[\Big[A_{ij}W_{ij}^{1}N_{i}(t,d) + B_{ij}(d)W_{ij}^{2}N_{i}(t,d)\Big]\Big| X_i, X_j\Big]}
	\label{estimator_2}
	\end{split}
	\end{equation}
	
	We simplify the inner terms in the numerator and denominator in the expression in \eqref{estimator_2} below.

	\begin{equation}
	\begin{split}
	&E[A_{ij}W_{ij}^{1}N_{i}(t,d)| X_i, X_j] = \int_{0}^{t} G(s|X_j) S(s|X_j) G(s-|X_j) W_{ij}^{1} dF_d(s|X_j) =  \int_{0}^{t} S(s|X_j)   dF_d(s|X_j)
	\end{split}
	\label{eqn2_prop2}
	\end{equation}
	
	\begin{equation}
	\begin{split}
	&E[B_{ij}W_{ij}^{2}N_{i}(t,d)| X_i, X_j] = \int_{0}^{t} \int_{0}^{s} G(s|X_j) S(s|X_j) G(s|X_j) W_{ij}^{1} dF_d(s|X_j) =  \int_{0}^{t} F_{d^{c}}(s|X_j)   dF_d(s|X_j)
	\end{split}
	\label{eqn3_prop2}
	\end{equation} 
In \eqref{eqn3_prop2}, $F_{d^{c}}$ is the CIF corresponding to the event other than $d$. 
	Substituting the above equations \eqref{eqn2_prop2}  \eqref{eqn3_prop2} in equation \eqref{estimator_2} we obtain 
	
		\begin{equation}
		\begin{split}
		\hat{\mathcal{JC}}_{wtd}(t) \rightarrow 	\sum_{d}\frac{E_{X_i, X_j}\Big[\int_{0}^{t}(1-F_d(s|X_j))dF_{d}(s|X_i)I(M(X_i,t, d)>M(X_j,t,d)\; \& \; M_c(X_i,t) = d )\Big]}{\sum_{d}E_{X_i, X_j}\Big[\int_{0}^{t}(1-F_d(s|X_j))dF_{d}(s|X_i)\Big]}	 
		\label{estimator_3}
		\end{split}
		\end{equation}
	
	  Since the expression in \eqref{estimator_3}, \eqref{simplify_JC1} are the same we conclude that the weighted estimator is consistent.

	\section*{References}

	\small
	
	[1] Wolbers, M., Koller, M. T., Stel, V. S., Schaer, B., Jager, K. J., Leffondre, K., \& Heinze, G. (2014). Competing risks analyses: objectives and approaches. European heart journal, 35(42), 2936-2941.

	[2] Jacobs, J., \& Fisher, P. (2013). Polypharmacy, multimorbidity and the value of integrative medicine in public health. European Journal of Integrative Medicine, 5(1), 4-7.
	
	[3] Wolbers, M., Koller, M. T., Witteman, J. C., \& Steyerberg, E. W. (2009). Prognostic models with competing risks: methods and application to coronary risk prediction. Epidemiology, 20(4), 555-56
	
	[4] Pefoyo, A. J. K., Bronskill, S. E., Gruneir, A., Calzavara, A., Thavorn, K., Petrosyan, Y., ... \& Wodchis, W. P. (2015). The increasing burden and complexity of multimorbidity. BMC Public Health, 15(1), 415.
	
	[5] Beyersmann, J., \& Schumacher, M. (2008). Time-dependent covariates in the proportional subdistribution hazards model for competing risks. Biostatistics, 9(4), 765-776.
	
	[6] Wolbers, M., Blanche, P., Koller, M. T., Witteman, J. C., \& Gerds, T. A. (2014). Concordance for prognostic models with competing risks. Biostatistics, 15(3), 526-539.
	
	[7] Crowder, M. J. (2001). Classical competing risks. CRC Press.
	
	[8] Daskivich, T. J., Chamie, K., Kwan, L., Labo, J., Dash, A., Greenfield, S., \& Litwin, M. S. (2011). Comorbidity and competing risks for mortality in men with prostate cancer. Cancer, 117(20), 4642-4650.
	
	[9] Murray, K., \& Conner, M. M. (2009). Methods to quantify variable importance: implications for the analysis of noisy ecological data. Ecology, 90(2), 348-355.
	
	[10] Thompson, B. (1995). Stepwise regression and stepwise discriminant analysis need not apply here: A guidelines editorial.
	
	[11] Puddu, P. E., Piras, P., \& Menotti, A. (2016). Competing risks and lifetime coronary heart disease incidence during 50 years of follow-up. International journal of cardiology, 219, 79-83.

	[12]  Atkins, D., Psaty, B. M., Koepsell, T. D., Longstreth, W. T., \& Larson, E. B. (1993). Cholesterol reduction and the risk for stroke in men: a meta-analysis of randomized, controlled trials. Annals of internal medicine, 119(2), 136-145.
	
	[13] Fine, J. P., \& Gray, R. J. (1999). A proportional hazards model for the subdistribution of a competing risk. Journal of the American statistical association, 94(446), 496-509.
	
%	[14] Ahuja, K., \& van der Schaar, M. (2018). Supplementary Materials for Joint Concordance Index. \url{https://drive.google.com/file/d/1TjcQxD7E1wCHmNQ-mGqCZFuL1IiW_Oet/view}
%	
	[14] Yoo, W., \& Coughlin, S. S. (2019). Surveillance, Epidemiology, and End Results (SEER) data for monitoring cancer trends. Journal of the Georgia Public Health Association.
\end{document}